\documentclass[conference]{IEEEtran}
\pdfoutput=1
\usepackage{amsmath}
\usepackage{amssymb}
\usepackage{graphicx}
\usepackage{subfigure}
\usepackage{enumerate}
\usepackage{amsthm}
\usepackage{comment}
\usepackage{mathtools}
\usepackage{scrextend} 

\newtheorem{thm}{Theorem}
\newtheorem{lemma}[thm]{Lemma}

\newtheorem{proposition}[thm]{Proposition}
\newtheorem{corollary}[thm]{Corollary}

\newcommand\numberthis{\addtocounter{equation}{1}\tag{\theequation}} 

\begin{document}


\title{Active Search with a Cost for Switching Actions}

\author{\IEEEauthorblockN{Nidhin Koshy Vaidhiyan} \IEEEauthorblockA{nidhinkv@ece.iisc.ernet.in \\Department of ECE\\Indian Institute of Science \\Bangalore 560012, India}  \and \IEEEauthorblockN{Rajesh Sundaresan} \IEEEauthorblockA{rajeshs@ece.iisc.ernet.in \\Department of ECE\\Indian Institute of Science \\Bangalore 560012, India}}

 \maketitle

\begin{abstract}
  Active Sequential Hypothesis Testing (ASHT) is an extension of the classical sequential hypothesis testing problem with controls.  Chernoff \cite{ref:195909AMS_Che} proposed a policy called {\it Procedure A} and showed its asymptotic optimality as the  cost of sampling was driven to zero. In this paper we study a further extension where we introduce costs for switching of actions. We show that a modification of Chernoff's {\it Procedure A}, one that we call {\it Sluggish Procedure A}, is asymptotically optimal even with switching costs. The growth rate of the total cost, as the probability of false detection is driven to zero, and as a switching parameter of the {\it Sluggish Procedure A} is driven down to zero, is the same as that without switching costs. 
\end{abstract}

\section{Introduction}

 Active Sequential Hypothesis Testing (ASHT) is a generalization of the classical sequential hypothesis testing problem where, at each observation instant, the decision maker has a choice that controls the type or quality of the observation. For example, in a cognitive radio setting, at each observation instant, the decision maker must select exactly one frequency band, of the several available, for observation. Another example is visual search where, at each time instant, one can focus only on a small subset of the entire visual field, and one must choose this subset for information gathering. ASHT can be used as a modeling tool for many other applications apart from  visual search and  cognitive radio, such as anomaly detection, medical diagnostics, etc.

Chernoff \cite{ref:195909AMS_Che} studied ASHT in the context of designing optimal experiments. His performance metric was the total cost of sampling, which is propotional to delay, plus a penalty for false detection. Chernoff proposed a policy, the so-called {\it Procedure A},  and showed its asymptotic optimality as the cost of sampling went to zero. {\it Procedure A} maintains a posterior distribution on the set of hypotheses and, at each instant, selects actions according to the hypothesis with the highest posterior probability.

In this paper we study a further extension of ASHT where, in addition to the average decision delay, we also penalize switching of actions. The current extension is motivated by visual search where a switch in action implies a change in the location of one's focus and a fast movement of the eyes (a saccade), which has an associated biological cost that translates to a delay cost.

We propose a modified {\it Procedure A} where the next action depends on the current posterior and the previous action, whereas in {\it Procedure A} the next action depends only on the current posterior. The modification is simple: at a given decision instant, if an independently generated Bernoulli random variable turns up ``$1$'', then the next action is taken as per {\it Procedure A}, else the current action is continued. We call this the {\it Sluggish Procedure A}. 

There has been a flurry of recent activity extending Chernoff's work in other directions. In a series of works, Naghshvar and Javidi \cite{ ref:201312TAS_NagJav, ref:201006ISIT_NagJav, ref:201010Alt_NagJav, ref:201108ISIT_NagJav, ref:201310JSP_NagJav} studied ASHT from a Bayesian cost minimization perspective. The total cost was the sum of decision delay and a penalty for false detection. They proposed policies, similar to Chernoff's {\it Procedure A}, identified bounds on the total cost, and established their proposed policies' asymptotic optimality in the same asymptotic regime as Chernoff's\footnote{They also consider the asymptotics where the number of hypotheses is large. This is not of direct relevance to our study.}. Nitinawarat et al. \cite{ref:201310ITAC_NitinVeeravalli} studied active hypothesis testing in fixed sample size and in sequential settings. They also minimize decision delay subject to a constraint on the conditional probability of false detection. When these conditional probabilities of false detection are driven to zero, the resulting asymptotic regime is the same as Chernoff's. In this asymptotic regime, they obtained results similar to those of Chernoff's but under milder assumptions. They also prove a stronger asymptotic result based on the ``risk associated with a decision''. Nitinawarat and Veeravalli \cite{2013arXiv1310.1844_NitinawaratVeeravalli} extended ASHT to Markovian observations and non-uniform costs on actions. Recently Cohen and Zhao \cite{ref:CohenZhao_AnomalyDetection_TIT_Mar2015} studied ASHT from an anomaly detection perspective. They showed that, in their setting, a simple deterministic policy was optimal. This is in contrast to random policies advocated in the other works. None of the above works consider costs associated with a change in action.

\paragraph*{Our contribution}
We show that the aforementioned modification of Chernoff's {\it Procedure A}, the {\it Sluggish Procedure A}, is asymptotically optimal even with switching costs. Further, we show that the growth rate of the total cost, as the probability of false detection is driven to zero, and as a switching parameter of the {\it Sluggish Procedure A} is driven to zero, is the same as that without switching costs.

\section{The ASHT Abstraction}
\label{sec:ASHT problem}
In this section, we describe our mathematical model for ASHT and collect all the  relevant theoretical results.

\subsection{The ASHT Model}
\subsubsection{The model description}
\label{sec:basic notation}
Let us begin by setting up some notation.

Let $H_{i}$, $i = 1,2, \ldots,M$ denote the $M$ hypotheses of which exactly one, denoted $H$, holds true. We do not assume a prior on the hypotheses. Let $\mathcal{A}$ be the set of all possible actions which we take as finite: $|\mathcal{A}| = K < \infty$. Let $\mathcal{X}$ be the observation space. Let $(X_{n})_{n \ge 1}$ and $(A_{n})_{n \ge 1}$ denote the observation process and the control process respectively.  We write $X^n$ for $(X_1, \ldots, X_n)$ and similarly $A^n$ for $(A_1, \ldots, A_n)$. We also write $\mathcal{P}(\mathcal{A})$ for the set of probability distributions on $\mathcal{A}$.

A policy $\pi$ is a sequence of action plans that at time $n$ looks at the history ${X}^{n-1},{A}^{n-1}$ and prescribes a composite action that is either $(stop, \delta)$ or $(continue, \lambda)$ as explained next. If the composite action is $(stop, \delta)$, then the controller stops taking further samples (or retires) and indicates $\delta$ as its decision on the hypothesis; $\delta \in \{1, 2, \ldots, M\}$. If the composite action is $(continue, \lambda)$, the controller picks the next action $A_{n}$ according to the distribution $\lambda \in \mathcal{P}(\mathcal{A})$. Let $\tau(\pi)$ be the stopping time $$\tau(\pi) := \inf\{n \geq 1 | A_{n} = (stop,\cdot)\}.$$

Consider a policy $\pi$. Conditioned on action $A_{n}$ and  the true hypothesis $H$, we assume that $X_{n}$ is conditionally independent of previous actions ${{A}^{n-1}}=(A_{1}, A_{2},\dots, A_{n-1})$, previous observations  ${{X}^{n-1}}=(X_{1}, X_{2}, \dots, X_{n-1})$, and the policy. Let $q_{i}^{a}$ be the conditional probability density function, with respect to some reference measure $\mu$, of the observation $X_{n}$ under action $a$ when $H = H_{i}$. Let $D(q_{i}^{a}\Vert q_{j}^{a})$ denote the relative entropy\footnote{By an abuse of notation, we use the densities of the probability measures as the arguments of the relative entropy function.} between the conditional probability measures associated with the observations under hypothesis $H_{i}$ and under hypothesis $H_{j}$, upon action $a$. Denote by  \textsf{unif($\mathcal{A}$)} the uniform distribution on $\mathcal{A}$. Let $q_{i}^{\pi}({x^{n}},{a^{n}})$ be the probability density function of observations and actions $({x^{n}},{a^{n}})$ till time $n$, with respect to the common reference measure $\mu^{\otimes n} \times \textsf{unif($\mathcal{A}$)}^{\otimes n}$. Let $Z_{i}^{\pi}(n)$ denote the log-likelihood process of hypothesis $H_{i}$, i.e.,
\begin{align}
 \label{eqn:general LLP}   Z_{i}^{\pi}(n) &= \log {q_{i}^{\pi}\left({{X}^{n}},{{A}^{n}} \right)}.
\end{align}
 Going forward, for ease of notation, we drop the superscript $\pi$ while describing $q_{i}^{\pi}$, $Z_{i}^{\pi}$, and other variables, but their dependence on the underlying policy should be kept in mind, and the policy under consideration will be clear from the context.
Define $Z(n) = (Z_{1}(n), Z_{2}(n), \ldots, Z_{M}(n))$.
Let $Z_{ij}(n)$ denote the log-likelihood ratio (LLR) process of $H_{i}$ with respect to $H_{j}$, i.e.,
\begin{align*}
  Z_{ij}(n) &= Z_{i}(n) - Z_{j}(n)\\
  \nonumber &= \log \frac{q_{i}\left({{X}^{n}},{{A}^{n}} \right)}{q_{j}\left({{X}^{n}},{{A}^{n}} \right)}\\
  \nonumber &=  \sum_{l=1}^{n}\log\frac{ q_{i}^{A_{l}}\left(X_{l} \right)}{  q_{j}^{A_{l}}\left(X_{l} \right)}.
\end{align*}

Let $E_{i}$ denote the conditional expectation and let $P_{i}$ denote the conditional probability measure under $H = H_{i}$. (More formally, these should be represented $E_{i}^{\pi}$ and $P_{i}^{\pi}$. But as done above, we omit the superscript $\pi$.)

Given an error tolerance vector $\alpha = (\alpha_{1}, \alpha_{2}, \ldots, \alpha_{M})$ with $0<\alpha_{i}<1$, let $\Pi(\alpha)$ be the set of policies
\begin{align*}
 \Pi(\alpha)= \left\{\pi: P_{i}(d \ne i) \le \alpha_{i}, \; \forall \; i \right\}.
\end{align*}
These are policies that meet a specified tolerance for the conditional probability of false detection. We define $\Vert \alpha \Vert := \max_i \alpha_i$.

We define $\lambda_{i}$ to be the best mixed action that guards $H_{i}$ against its nearest alternative, i.e., $\lambda_{i} \in \mathcal{P}(\mathcal{A})$ such that
  \begin{align}
   \label{eqn:optimal_lambda}
   \lambda_{i} := \arg \max_{\lambda \in \mathcal{P}(\mathcal{A})} \left[ \min_{j \ne i} \sum_{a  \in \mathcal{A}} \lambda(a) D(q_{i}^{a}\Vert q_{j}^{a})\right]. 
  \end{align}
If there are several maximizers, pick one arbitrarily. Further, define
 \begin{align}
\label{eqn:D_i}
  D_{i} := \max_{\lambda \in \mathcal{P}(\mathcal{A})} \left[\min_{j \ne i} \sum_{a \in \mathcal{A}} \lambda({a}) D\left(q_{i}^{a}\Vert q_{j}^{a}\right)\right].
\end{align}
Let $\mathcal{A}_{ij} = \{a \in \mathcal{A}: D(q_{i}^{a}\Vert q_{j}^{a})>0\}$, the set of all actions that can differentiate hypothesis $H_{i}$ from hypothesis $H_{j}$. From well known properties of relative entropy, we obtain $\mathcal{A}_{ij} = \mathcal{A}_{ji}$. 

\subsubsection{Assumptions}
Throughout, we make the following assumptions.
\begin{enumerate}
\item [(I)\;\;  ] $E_{i}\left[\left(\log \frac{q_{i}^{a}(X)}{q_{j}^{a}(X)} \right)^{2}\right] < \infty$ $\forall$ $i,j,a$.\\
\item [(IIa)] \label{assumpIIa} $\mathcal{A}_{ij} \ne \varnothing \quad \forall i,j \quad i \ne j.$
\item [(IIb)] \label{assumpIIb} $\beta := \min_{i \ne j,k} \sum_{a \in \mathcal{A}_{ij}} \lambda_{k}(a) > 0$.
\end{enumerate}

Assumption (I) implies that $D(q^a_i || q^a_j) < \infty$, which in turn ensures that no single observation can result in a reliable decision. Assumption (I) is used in proving the lower bound on the expected number of samples needed to satisfy the tolerance criterion. This is also assumed by Chernoff \cite{ref:195909AMS_Che} and Nitinawarat et al. \cite{ref:201310ITAC_NitinVeeravalli}.

Assumption (IIa) ensures that for any distinct $i$ and $j$, there is at least one control that can help distinguish the hypotheses $H_{i}$ from $H_{j}$. If $\mathcal{A}_{ij} = \emptyset$ for some $i$ and $j$, it will be impossible to distinguish them from each other. Assumption (IIb) is a stronger assumption than, and implies, Assumption (IIa). Assumption (IIb) ensures that if actions are taken according to any of the $\lambda_{k}$ in (\ref{eqn:optimal_lambda}) then, for any two hypotheses $H_{i}$ and $H_{j}$, there is a positive probability of choosing an action that can discriminate them. We shall use Assumption (IIb) in the achievability proofs of our policies. It allows for easier proofs for our policies, and makes the presentation simpler. However one can work with Assumption (IIa) as well, and construct asymptotically optimal policies, with minor modifications to our policies. We will describe the modifications later in this section.

\subsubsection{Switching cost and total cost}
The costs are as follows.
\paragraph*{Switching Cost}
Let $g(a,a')$ denote the cost of switching from action $a$ to action $a'$. We assume $$g(a,a') \ge 0 \quad \forall a,a' \in \mathcal{A} \quad \text{and} \quad g({a,a}) = 0.$$
Define  $g_{\max} = \max_{a,a'} g(a,a')$. We also assume $g_{\max} < \infty$.
\paragraph*{Total cost}
\label{sec:performance criterion}
For a policy $\pi \in \Pi(\alpha)$, the total cost $C(\pi)$ is taken to be the sum of the stopping time (delay) and the net switching cost, i.e.,  $$C(\pi) := \tau(\pi)+ \sum_{l=1}^{\tau(\pi)-1} g({A_{l},A_{l+1}}).$$ 

\subsubsection{Asymptotics}
We shall be interested in the asymptotics of the minimum expected total cost $E_{i}[C(\pi)]$, minimized over policies in $\Pi(\alpha)$, as $||\alpha|| \rightarrow 0$. Note that there are $M$ such conditional expected total costs, one for each hypothesis.

\subsection{Results on the ASHT Model}
We collect all the main results in this section. We first identify a lower bound.
\subsubsection{The converse - Lower bound}
\label{sec:lower bounds}
The following proposition gives a lower bound for the expected conditional stopping time, given hypothesis $H = H_{i}$, for all policies belonging to $\Pi(\alpha)$. 
\begin{proposition}
 \label{proposition:proposition1}
Assume (I). For each $i$, we have
 \begin{align}
\label{eqn:lower bound} \lim _{\Vert \alpha \Vert \rightarrow 0} \inf_{\pi \in \Pi(\alpha)} \frac{E_{i} [\tau(\pi)]}{\vert \log \Vert \alpha \Vert \vert} \ge \frac{1}{D_{i}}, 
 \end{align}
where $D_{i}$ is given in (\ref{eqn:D_i}).
 \end{proposition}
\begin{IEEEproof}
 Since only expected time to stop is considered, proof of \cite[Th. 2, p. 766]{ref:195909AMS_Che} applies.
\end{IEEEproof}
\vspace*{0.1 in}
We then have the following corollary.
\begin{corollary}
Assume (I). For each $i$, we have
{\small
\label{cor:total_cost_lower_bound}
 \begin{align}
\label{eqn:total_cost_lower_bound} \lim_{\Vert \alpha \Vert \rightarrow 0} \inf_{\pi \in \Pi(\alpha)} \frac{E_{i}[ C(\pi)]}{|\log \Vert \alpha \Vert |} \ge \frac{1}{D_{i}}.
 \end{align}
 }
\end{corollary}
\begin{IEEEproof}
 With switching costs added, we have $C(\pi) \ge \tau(\pi)$, and the corollary follows from Propostion \ref{proposition:proposition1}. 
\end{IEEEproof}
\vspace*{0.1 in}

\subsubsection{Achievability -  A modification to Chernoff's Procedure A}
\label{sec:upper bounds}
Chernoff \cite{ref:195909AMS_Che} proposed a policy termed {\it Procedure A} and showed that it has asymptotically optimal expected decision delay. We now describe {\it Procedure A}.

\begin{addmargin}[2em]{2em}
{\it Policy {\it Procedure A:}} $\pi_{PA}(L)$ \\Fix $L > 0 $. \\
At time $n$:
\begin{itemize}
\item Let $\theta(n) = \arg\max_{i} Z_{i}(n)$. Ties are resolved uniformly at random.
\item If $ Z_{\theta(n),j}(n) < \log{((M-1)L)}$ for some $j \ne \theta(n)$ then $A_{n+1}$ is chosen according to $\lambda_{\theta(n)}$, i.e.,
\begin{align}
\label{eqn:probability of action selection}
 \Pr(A_{n+1} = a) = \lambda_{\theta(n)}(a)
\end{align}
\item If $Z_{\theta(n),j}(n) \ge \log{((M-1)L)}$ for all $j \ne \theta(n)$ then the test retires and declares $H_{\theta(n)}$ as the true hypothesis.
\end{itemize}
\end{addmargin}
We now describe a modified policy that can be made as close as one wishes to being asymptotically optimal in the presence of switching costs. We introduce a switching parameter $\eta, \: 0 < \eta \le 1 $, which determines the maximum transition rate out of a given action. When $\eta = 1$, we will have the original {\it Procedure A}. When $\eta$ approaches zero, the rate of jumping out of the current action approaches zero.
\begin{addmargin}[2em]{2em}
{\it Policy {\it Sluggish Procedure A:}} $\pi_{SA}(L,\eta)$ \\Fix $L > 0 , \: 0 < \eta \le 1$. \\
At time $n$:
\begin{itemize}
\item Let $\theta (n) = \arg\max_{i} Z_{i}(n)$. Ties are resolved uniformly at random.
\item If $ Z_{\theta(n),j}(n) < \log({(M-1)L})$ for some $j \ne \theta(n)$ then $A_{n+1}$ is chosen as follows.
\begin{itemize}
 \item Generate $U_{n+1}$, a Bernoulli($\eta$) random variable, independent of all other random variables.
 \item If $U_{n+1} = 0$, then $A_{n+1} = A_{n}$.
 \item If $U_{n+1} = 1$, then generate $A_{n+1}$ according to distribution $\lambda_{\theta(n)}$.
\end{itemize}

\item If $Z_{\theta(n),j}(n) \ge \log{(M-1)L}$, for all $j \ne \theta(n)$, then the test retires and declares $H_{\theta(n)}$ as the true hypothesis. 
\end{itemize}
\end{addmargin}
\vspace*{0.1in}

We also consider two variants of $\pi_{SA}(L,\eta)$ which are useful in the analysis.
\begin{itemize}
 \item {\it Policy} $\pi_{SA}^{i}(L,\eta)$: This is the same as $\pi_{SA}(L,\eta)$, but stops only at decision $i$ when $\min_{j \ne i} Z_{ij}(n) \ge \log(L (M-1))$.
 \item {\it Policy} $\tilde{\pi}_{SA}(\eta)$: This is the same as $\pi_{SA}(L,\eta)$, but never stops, and hence $L$ is irrelevant.
\end{itemize}
Under a fixed hypothesis $H = H_{i}$, and the triplet of policies $(\pi_{SA}(L,\eta), \pi_{SA}^{i}(L,\eta), \tilde{\pi}_{SA}(\eta))$, it is easily seen that there is a common underlying probability measure with respect to which the processes $(X_n, A_n)_{n \ge 1}$ associated with the three policies are naturally coupled, with only the stopping times being different. Under this coupling, the following are true:
\begin{align*}
 \tau(\pi_{SA}^{i}(L,\eta)) & \ge \tau(\pi_{SA}(L,\eta)),\\
 \{\tau (\pi_{SA}(L,\eta)) > n\} &\subset \{\tau(\pi_{SA}^{i}(L,\eta)) > n\}\\
&\subset \{\min_{j \ne i} Z_{ij}(n) < \log(L (M-1))\}.
 \end{align*}

Policy $\pi_{SA}(L,\eta)$ is designed to stop only when the posteriors suggest a reliable decision. This is formalized now.
\begin{proposition}
 \label{prop:probability of wrong detection}
Assume (I) and (IIb). For Policy ${\pi_{SA}(L,\eta)}$, the conditional probability of error under hypothesis $H_{i}$ is upper bounded by
\begin{align}
 P_{i}(d \ne i) \le \frac{1}{L}.
\end{align}
\end{proposition}

See Appendix \ref{appendix:proof of probability of wring detection} for a proof. As a consequence we have ${\pi_{SA}}(L,\eta) \in \Pi(\alpha)$ if $\alpha_{i} \ge \frac{1}{L} \quad \forall i$.
\vspace*{0.1 in}
We now state the time-delay performance of the policy $\pi_{SA}(L,\eta)$. 
\begin{thm}
 Assume (I) and (IIb). Consider the policy $\pi_{SA}(L, \eta)$. The expected time to make a decision, for each $i$, satisfies
\label{theorem: upper bound on stopping time of SA}
\begin{align}
\label{eqn:upper_bound_sandwich_argument}
\lim_{L \rightarrow \infty} \frac{E_{i}\left[\tau(\pi_{SA}(L,\eta))\right]}{\log L} \le \frac{1}{D_{i}}.
\end{align}
\end{thm}

See Appendix \ref{ref:lemma:achievability proof} for a detailed proof. This result will be crucial because the policy $\pi_{SA}(L, \eta)$, despite its sluggishness induced by $\eta$, remains asymptotically optimal when only the stopping time $\tau(\pi_{SA}(L,\eta))$ is considered as cost. We now leverage this to show that, if $\eta$ is sufficiently small, $\pi_{SA}(L,\eta)$ is near optimal when switching costs are also taken into account.

\begin{proposition}
 \label{prop:achievability}
 Assume (I) and (IIb). Consider the policy $\pi_{SA}(L, \eta)$. We then have, for each $i$,
 \begin{align}
 \label{eqn:achievability}
  \lim _{L \rightarrow \infty} E_{i}\left[\frac{C(\pi_{SA}(L,\eta))}{\log L}\right] & \le \frac{1}{D_{i}} + \frac{g_{\max}\eta }{D_{i}}.
 \end{align}
\end{proposition}

\begin{IEEEproof}
{We can write the following chain of inequalities.
\small
\begin{align}
\nonumber E_{i} &\left[C(\pi_{SA}(L,\eta))\right]\\
\nonumber  &= E_{i}\left[\tau(\pi_{SA}(L,\eta))+\sum_{l=1}^{\tau(\pi_{SA}(L,\eta))-1} g(A_{l},A_{l+1})\right]\\
\nonumber & \le E_{i}\left[\tau(\pi_{SA}(L,\eta))\right]+g_{\max}E_{i}\left[\sum_{l=1}^{\tau(\pi_{SA}(L,\eta))-1}1_{\{A_{l}\ne A_{l+1}\}}\right]\\
\nonumber & \le E_{i}\left[\tau(\pi_{SA}(L,\eta))\right]+g_{\max}E_{i}\left[\sum_{l=1}^{\tau(\pi_{SA}(L,\eta))-1}U_{l+1}\right]\\
\nonumber & = E_{i}\left[\tau(\pi_{SA}(L,\eta))\right]+g_{\max}\eta E_{i}\left[\tau(\pi_{SA}(L,\eta))-1\right]\\
\label{eqn:cost_bound}& \le E_{i}\left[\tau(\pi_{SA}(L,\eta))\right](1+g_{\max}\eta).
\end{align}
}
In the above chain, the penultimate equality holds because of Wald's equation \cite{ref:wald1944}. Dividing by $\log L$, letting $L \rightarrow \infty$, and using Theorem \ref{theorem: upper bound on stopping time of SA}, we see that (\ref{eqn:achievability}) holds.
\end{IEEEproof}
\subsubsection{Asymptotic optimality}
Proposition \ref{proposition:proposition1} and Proposition \ref{prop:achievability} show that, when the conditional probability of false detection is driven to zero, the proposed policy $\pi_{SA}(L,\eta)$ has nearly the same growth rate for cost as an asymptotically optimal policy without switching costs. We now make the above statement precise.  The parameter $\eta$ should be suitably chosen to get sufficiently close to asymptotic optimality.

\begin{thm}
\label{thm:asymptotic optimality of sluggish procedure A}
 Assume (I) and (IIb). Consider a sequence of vectors $(\alpha^{(n)})_{n \ge 1}$, where  $\alpha^{(n)}$ is the $n^{th}$ tolerance vector, such that  $\lim_{n \rightarrow \infty} \Vert \alpha^{(n)} \Vert = 0$ and
 \begin{align}
 \label{eqn:condition on max and min of tolerance vector}
   \lim_{n \rightarrow \infty} \frac{\Vert \alpha^{(n)} \Vert}{\min_{k}\alpha^{(n)}_{k}} < B
 \end{align} for some $B$. Then, the sequence of policies $\pi_{SA}(L_n,\eta)$ with $\log L_n = -\log \min_{k}\alpha^{(n)}_{k}$ belongs to $\Pi(\alpha^{(n)})$. Furthermore, for each $i$,
{\small
\begin{align}
\label{eqn:asymptotic optimality}
 \lim_{n \uparrow \infty}\inf_{\pi \in \Pi(\alpha^{(n)})} \frac{E_{i}\left[C(\pi)\right]}{\log L_n} = \lim_{\eta \downarrow 0}\lim_{n \uparrow \infty} \frac{E_{i}\left[C(\pi_{SA}(L_n,\eta))\right]}{\log L_n} = \frac{1}{D_{i}}.
\end{align}
}
\end{thm}

\begin{IEEEproof}
 The fact that  $\pi_{SA}(L_n,\eta) \in \Pi(\alpha^{(n)})$ is evident from Proposition \ref{prop:probability of wrong detection}, and $\frac{1}{L_{n}} \le \alpha^{n}_{k}, \: k = 1,2, \cdots, n$. We then have the following chain of inequalities:
\begin{align*}
 \frac{1}{D_{i}} &\le  \lim_{n \uparrow \infty}\inf_{\pi \in \Pi(\alpha^{(n)})} \frac{E_{i}\left[C(\pi)\right]}{|\log \Vert \alpha^{(n)} \Vert|}\\
 & = \lim_{n \uparrow \infty}\inf_{\pi \in \Pi(\alpha^{(n)})} \frac{E_{i}\left[C(\pi)\right]}{\log L_n}\\
 & \le \lim_{\eta \downarrow 0}\lim_{n \uparrow \infty} \frac{E_{i}\left[C(\pi_{SA}(L_n,\eta))\right]}{\log L_n}\\
 & \le \frac{1}{D_{i}}.
\end{align*}
The first inequality follows from Proposition \ref{proposition:proposition1}. The next equality follows from the fact that $\lim_{n \rightarrow \infty}\frac{|\log \Vert \alpha^{(n)} \Vert |}{\log L_{n}} = 1$, which in turn is true due to the assumption (\ref{eqn:condition on max and min of tolerance vector}). The third inequality follows because $\pi_{SA}(L_{n},\eta)$ is one specific policy in $\Pi(\alpha^{n})$. The last inequality follows from Proposition \ref{prop:achievability} after letting $\eta \downarrow 0$. Consequently, all inequalities must be equalities.
\end{IEEEproof}

\subsection{Discussion on Assumption (IIb)}
\label{sec:Assumption (II) discussion}
Chernoff's proof of the asymptotic optimality of {\it Procedure A} was proved under a stronger assumption than Assumption (IIb), namely, Chernoff required 
\begin{align}
\label{eqn:Chernoff's assumption}
D(q_{i}^{a} \Vert q_{j}^{a}) > 0  \quad \forall a \text{ and for all pairs $i \ne j$}. 
\end{align}
 Assumption (IIb) ensures that, at all times, and for any pair of hypotheses $i$ and $j$, $i \ne j$, there is a positive probability of choosing an action that can distinguish the two hypotheses. This suffices for Chernoff's proofs to go through. Specifically, we shall use Assumption (IIb) to prove the exponential decay result in Proposition \ref{prop:exponential decay LLR process}. Nitinawarat et al.  \cite{ref:201310ITAC_NitinVeeravalli} proposed a modified {\it Procedure A} that sampled actions randomly at intervals $\lceil{{\nu}^{l}\rceil}_{l \ge 1}, \; \nu > 1$, and showed that their proposed policy is asymptotically optimal under the weaker Assumption (IIa). The random sampling enabled them to obtain a polynomial decay counterpart of Proposition \ref{prop:exponential decay LLR process} of Appendix. Recently, Cohen and Zhao \cite{ref:CohenZhao_AnomalyDetection_TIT_Mar2015} claimed the asymptotic optimality of {\it Procedure A} under the weaker Assumption (IIa) for an active anomaly detection problem, which is a specific ASHT problem. We conjecture that Chernoff's {\it Procedure A} is asymptotically optimal under the weaker Assumption (IIa) for all ASHT problems. A proof of this claim has remained elusive. Nevertheless, policies whose performances are provably arbitrarily close to the optimum can be designed. We make the above claim precise in the next proposition.

\begin{proposition}
 \label{prop: Near optimal provable policies}
 Assume (I) and (IIa). Fix $\epsilon > 0$. Then there exists a sequence of policies $\{\pi_{\epsilon}(L)\}$ that satisfies $\pi_{\epsilon}(L) \in \Pi(\frac{1}{L},\frac{1}{L},\cdots,\frac{1}{L})$ and 
 \begin{align}
  \lim _{L \rightarrow \infty} E_{i}\left[\frac{\tau(\pi_{\epsilon}(L))}{\log L}\right] & \le \frac{1}{(1-\epsilon)D_{i}}.
 \end{align}
\end{proposition}
We omit the proof because the needed modifications to the proof of Theorem \ref{theorem: upper bound on stopping time of SA} are straightforward. Policy $\{\pi_{\epsilon}(L)\}$ can be constructed as a variant of {\it Procedure A} that, at each instant $n$, chooses an action according to $\textsf{unif}(\mathcal{A})$ with probability $\epsilon$ or as per (\ref{eqn:probability of action selection}) with probability $(1-\epsilon)$. Thus, at the cost of a small penalty, we can design nearly asymptotically optimal policies under the weaker Assumption (IIa). A similar argument holds true with switching costs, just as Theorem \ref{theorem: upper bound on stopping time of SA} is extended in Theorem \ref{thm:asymptotic optimality of sluggish procedure A}, albeit with a corresponding but arbitrarily small increase in the total cost. Again, we omit the proof of this claim. Hence Assumption (IIa) suffices for the asymptotic growth rate to be $\frac{1}{D_{i}}$.

\section{Conclusion}
We studied active sequential hypothesis testing (ASHT) with switching costs. We proposed a modification to Chernoff's {\it Procedure A} that can be made to approach the asymptotic performance of {\it Procedure A}. The proposed algorithm merely slows down the switching of actions via an i.i.d. Bernoulli modulation process. The growth rate of total cost, as the probability of false detection is driven to zero, and as the switching parameter $\eta$ is driven to zero, is the same as that without switching costs.

\section*{Acknowledgments}
This work was supported in part by the University Grants Commission by Grant Part(2B) UGC-CAS-(Ph.IV) and in part by the Department of
Science and Technology.
\section*{Appendix}

\section{Properties of log-likelihood ratio processes under $\pi_{SA}(L,\eta)$}
\label{sec:app:PropertiesOfLLR}

We will now show some desirable properties of the log-likelihood ratio processes under the policy $\pi_{SA}(L, \eta)$. These properties are analogous to those of classical sequential hypothesis testing, but their analyses are more involved because actions introduce 1) dependency in the log-likelihood ratio increments, and 2) the  increments are no longer identically distributed. The properties we will establish will be useful in forthcoming proofs.

Define $\Delta Z_{ji}(n) = Z_{ji}(n)-Z_{ji}(n-1)$. We then have $ \Delta Z_{ji}(n) = -\Delta Z_{ij}(n)$. Here, $\Delta Z_{ji}(n)$ is the increment in the process associated with the log-likelihood ratio of $H_{j}$ with respect to $H_{i}$ at time $n$. We now show that under Assumptions (I) and (IIb), and under policy $\pi_{SA}(L, \eta)$, the log-likelihood ratio processes are well behaved in the following sense: the log-likelihood ratio of the true hypothesis $H_{i}$ with respect to any other hypothesis $H_{j}$ has a positive drift. This will be made precise in Proposition \ref{prop:exponential decay LLR process}. Towards that, we first establish the following lemmas.

\begin{lemma}
 \label{lemma:likelihood ratio property conditioned on actions}
Assume (I) and (IIb).  Fix $i$, $j$ such that $j \ne i$. Let $a \in \mathcal{A}_{ij}$. We then have, for all $0 <s <1$,
\begin{align}
 \rho_{ij}^{a}(s) := E_{i}\left[e^{s\Delta Z_{ji}(n)} \vert A_{n} = a\right] < 1 \;\;\; \hfill \forall n.
\end{align}
\end{lemma}

\begin{IEEEproof}
The following sequence of inequalities hold:
\begin{align}
 \nonumber E_{i} &\left[e^{s\Delta Z_{ji}(n)} \vert A_{n} = a\right]\\
 \nonumber &= \int_{x \in \mathcal{X}} \left(\frac{q_{j}^{a}(x)}{q_{i}^{a}(x)}\right)^{s} q_{i}^{a}(x) dx\\
 \nonumber &=  \int_{x \in \mathcal{X}} \left(q_{j}^{a}(x)\right)^{s} \left(q_{i}^{a}(x)\right)^{1-s} dx \\
& < \left( \int_{x \in \mathcal{X}} q_{j}^{a}(x) dx \right)^{s} \left( \int_{x \in \mathcal{X}} q_{j}^{a}(x) dx \right)^{1-s} \label{eqn:Holders ineq}\\
 \nonumber& = 1.
\end{align}
The strict inequality in (\ref{eqn:Holders ineq}) follows from H\"{o}lder's inequality and the fact that $a \in \mathcal{A}_{ij}$ implies $q_{i}^{a}$ and $q_{j}^{a}$ are not linearly related.
\end{IEEEproof}

The above result was obtained by conditioning on the action $A_{n}$ to lie in the desirable set $\mathcal{A}_{ij}$. The result is independent of the underlying policy, because when conditioned on the current action $A_{n}$, the observation is independent of the policy.

Recall that  $\tilde{\pi}_{SA}(\eta)$ is the non-stopping variant of ${\pi}_{SA}(L, \eta)$. Further, recall from Assumption (IIb) that we have $\beta =  \min \left\{ \sum_{a \in \mathcal{A}_{ij}} \lambda_{k}(a) ~|~ 1 \leq i, j, k \leq M, ~ i \ne j \right\} > 0$. Now we show that, under Assumption (IIb) and policy $\tilde{\pi}_{SA}(\eta)$, a similar result holds, but without conditioning on the action $A_{n}$.
First, let us define
\begin{align}
 \rho_{ij}(s) :=  \eta \beta \left(\max_{a \in \mathcal{A}_{ij}}\rho_{ij}^{a}(s)\right)+ (1-\eta \beta).
\end{align}
The fact that $\rho_{ij}(s) < 1$ is evident from Lemma \ref{lemma:likelihood ratio property conditioned on actions}.

\begin{lemma}
 \label{lemma:likelihood ratio property under policy}
 Assume (I) and (IIb). Consider the  policy $\tilde{\pi}_{SA}(\eta)$. Fix $i$. We then have, for all $0<s<1$,
\begin{align*}
 E_{i} &\left[e^{s\Delta Z_{ji}(n)} \vert X^{n-1}, A^{n-1}\right] \le \rho_{ij}(s) < 1 \; \; \; \hfill \forall n, \forall j \ne i.
\end{align*}
\end{lemma}

\begin{IEEEproof}
The following sequence of inequalities hold as described after the last inequality.
\begin{align}
\nonumber E_{i} &\left[e^{s\Delta Z_{ji}(n)} \vert X^{n-1},A^{n-1}\right]\\
\nonumber &= E_{i} \left[ E_{i} \left[ e^{s\Delta Z_{ji}(n)} \vert X^{n-1},A^{n-1}, A_{n} \right] \vert X^{n-1},A^{n-1} \right]\\
\label{eqn:lemma rho_ij(s) equation 3} &=  \sum_{a \in \mathcal{A}} P_{i}(A_{n} =a\vert X^{n-1} A^{n-1}) E_{i} \left[e^{s\Delta Z_{ji}(n)} \vert A_{n} =a \right]\\
\nonumber &\le P_{i}(A_{n} \in \mathcal{A}_{ij}\vert X^{n-1} A^{n-1}) \max_{a \in \mathcal{A}_{ij}} E_{i} \left[e^{s\Delta Z_{ji}(n)} \vert A_{n} =a\right]\\
\label{eqn:lemma rho_ij(s) equation 4}& \hspace{0.5cm}+ (1-P_{i}(A_{n} \in \mathcal{A}_{ij}\vert X^{n-1} A^{n-1}))\\
\nonumber & \le \eta \beta \left(\max_{a \in \mathcal{A}_{ij}}\rho_{ij}^{a}(s)\right)+ (1-\eta \beta)\\
& < 1.
\end{align}
Equality (\ref{eqn:lemma rho_ij(s) equation 3}) holds because conditioned on $A_{n} = a$, $\Delta Z_{ij}(n)$ is independent of the remaining history. Inequality (\ref{eqn:lemma rho_ij(s) equation 4}) holds because, when $a \notin \mathcal{A}_{ij}$, we have $\Delta Z_{ij}(n) \equiv 0$. The penultimate inequality is a consequence of the fact that, under $\pi_{SA}(L,\eta)$, one will choose an action $a \in \mathcal{A}_{ij}$ with probability at least $\eta \beta$.
\end{IEEEproof}

We now proceed to show  an inequality analogous to the Chernoff bound for the log-likelihood ratio. In classical sequential hypothesis testing, due to independence of samples across time, the expectation of the likelihood ratio can be split as the product of the expectation of the likelihood ratio increments, as follows:
\begin{align*}
 E_{i} \left[e^{sZ_{ji}(n)} \right] = \prod_{k=1}^{n} E_{i} \left[e^{s\Delta Z_{ji}(n)} \right].
\end{align*}
The same decomposition is not valid in ASHT because actions introduce dependency in the likelihood ratio increments across time. However, we can obtain an upper bound of the product form.

\begin{lemma}
\label{lemma:likelihood ratio decomposition}
 Assume (I) and (IIb). Consider policy $\tilde{\pi}_{SA}(\eta)$. Fix $i$. We then have, for all $0 <s<1$,
\begin{align*}
 E_{i} \left[e^{s Z_{ji}(n)}\right] \le (\rho_{ij}(s))^{n} \; \; \; \hfill \forall n, \forall j \ne i.
\end{align*}
\end{lemma}

\begin{IEEEproof}
Once again, we proceed through the chain of inequalities all of which are now self-evident:
\begin{align*}
 E_{i} &\left[e^{s Z_{ji}(n)} \right]\\
&= E_{i} \left[ E_{i} \left[ e^{s Z_{ji}(n-1)}e^{s\Delta Z_{ji}(n)} \vert X^{n-1},A^{n-1} \right] \right]\\
&=  E_{i} \left[e^{s Z_{ji}(n-1)} E_{i} \left[e^{s\Delta Z_{ji}(n)}\vert X^{n-1},A^{n-1} \right]  \right] \\
& = \rho_{ij}(s) E_{i} \left[e^{s Z_{ji}(n-1)}\right] \text{\hfill (from Lemma \ref{lemma:likelihood ratio property under policy})}\\
&\le (\rho_{ij}(s))^{n},
\end{align*}
where the last inequality follows by induction.
\end{IEEEproof}

We now show an exponential decay property of the log-likelihood process which primarily stems from the anticipated negative drift in $Z_{ji}(n)$ for $j\ne i$. Let us alert the reader that in the following Proposition we deal with $Z_{ij}(n) = -Z_{ji}(n)$.

\begin{proposition}
\label{prop:exponential decay LLR process}
 Assume (I) and (IIb). Consider policy $\tilde{\pi}_{SA}(\eta)$. Fix $i$. There exist constants $C_{K} > 0$ and $\gamma > 0$ such that
\begin{align}
\label{eqn:exp decay LLR ratio}
 P_{i}\left(\min_{j\ne i} Z_{ij}(n) \le K \right) < C_{K} e^{-\gamma n}.
\end{align}
$C_{K}$ is independent of $i$, but $\gamma$ may depend on $i$.
\end{proposition}

\begin{IEEEproof}
This follows from the previous lemmas via the following :
 \begin{align}
 \nonumber P_{i}\left(\min_{j\ne i} Z_{ij}(n) \le K \right) &= P_{i}\left(\max_{j\ne i} Z_{ji}(n) \ge -K \right)\\
 \label{eqn: exponential decay LLR process inequality 3} &\le \sum_{j \ne i} P_{i}\left(Z_{ji}(n) \ge -K \right)\\
 \label{eqn: exponential decay LLR process inequality 1} & \le \sum_{j \ne i} e^{sK}E_{i}\left[e^{sZ_{ji}(n)}\right]\\
 \label{eqn: exponential decay LLR process inequality 2} & \le e^{sK} \sum_{j \ne i} (\rho_{ij}(s))^{n}\\
 \nonumber & \le e^{sK} \cdot (M-1) \cdot \max_{j\ne i} (\rho_{ij}(s))^{n}\\
 \nonumber & = C_{K} e^{-\gamma n},
 \end{align}
where $\max_{j \ne i} \rho_{ij}(s) = e^{-\gamma}$, and $C_{K} = M e^{sK}$. The inequality in (\ref{eqn: exponential decay LLR process inequality 3}) is due to the union bound, the inequality in (\ref{eqn: exponential decay LLR process inequality 1}) is due to Chernoff's bound with $0<s<1$,  and the inequality in (\ref{eqn: exponential decay LLR process inequality 2}) is due to Lemma \ref{lemma:likelihood ratio decomposition}.
\end{IEEEproof}


 We now show that under the hypothesis $H = H_{i}$, the $\theta(n)$ process eventually settles at $i$.  Indeed we show something stronger. Let us define
\begin{align}
\label{eqn:Ti}
 T_{i} := \inf\{n:\theta(n') = i, \quad \forall n' \ge n\},
\end{align}
the time at which $\theta(n)$  meets its eventuality of settlement at $i$. This random variable has a tail that decays exponentially fast, as shown next.

\begin{lemma}
\label{lemma:exponential decay of Ti}
 Assume (I) and (IIb). Consider policy $\tilde{\pi}_{SA}(\eta)$. Fix $i$. Then there exist $C>0$ and $b >0$, both finite and possibly dependent on $i$, such that
\begin{align}
\label{eqn:exp decay Ti}
 P_{i}\left(T_{i} > n \right) < C e^{-bn}. \quad
\end{align}
\end{lemma}

\begin{IEEEproof}
By the union bound
 \begin{align*}
 P_{i}\left(T_{i} > n \right) & = P_{i}(\theta(n') \ne i \text{ for some $n' \ge n$})\\
 & \le  \sum_{n' \ge n}P_{i}\left(\theta(n') \ne i\right)\\
 & \le \sum_{n' \ge n}P_{i}\left(\min_{j \ne i} Z_{ij}(n') \le 0\right).
 \end{align*}
The assertion now follows from Proposition \ref{prop:exponential decay LLR process}.
\end{IEEEproof}


Thus far we have considered the policy $\tilde{\pi}_{SA}(\eta)$ which never stops. We now show that the policy $\pi_{SA}(L,\eta)$ stops in finite time.
\begin{proposition}
\label{prop:finite stopping time}
 Assume (I) and (IIb). Consider the policy $\pi_{SA}(L,\eta)$. Fix $i$. We then have
\begin{align*}
 P_{i}(\tau(\pi_{SA}(L,\eta)) < \infty) = 1.
\end{align*}
\end{proposition}
\begin{IEEEproof}
 We consider $\pi_{SA}^{i}(L,\eta)$  for analysis. Recall that $\tau(\pi_{SA}(L,\eta)) \le \tau(\pi_{SA}^{i}(L,\eta))$, and hence it is sufficient to show that
\begin{align}
\label{eqn:stopping time pi_i finite}
 P_{i}(\tau(\pi_{SA}^{i}(L,\eta) < \infty) = 1.
\end{align}
From Proposition \ref{prop:exponential decay LLR process}, we know that, for a suitable constant $\tilde{C}$, $$P_{i}\left(\min_{j\ne i} Z_{ij}(n) < \log(L(M-1)) \right) < \tilde{C} e^{-\gamma n}.$$ Since this bound is summable, by the Borel-Cantelli lemma, $$P_{i}\left(\min_{j\ne i} Z_{ij}(n) < \log(L(M-1)) \quad \text{infinitely often}\right) = 0,$$ which is stronger than the assertion (\ref{eqn:stopping time pi_i finite}).
\end{IEEEproof}
Propositions \ref{prop:exponential decay LLR process} and \ref{prop:finite stopping time} are the ones that will be used in the sequel.

\subsection{Proof of Proposition \ref{prop:probability of wrong detection}}
\label{appendix:proof of probability of wring detection}

The proof relies on a standard change of measure argument. Let $\Delta_{j}$ denote the event that the policy $\pi_{SA}(L,\eta)$ declares $H_{j}$ as the true hypothesis.
 \begin{align}
\nonumber  P_{i}(\delta \ne i) &= \sum_{j \ne i} P_{i}(\delta = j)+ P_{i}(\tau(\pi_{SA}(L,\eta)) = \infty)\\
\nonumber &= \sum_{j \ne i} \sum_{n>0}\int_{\omega^{n} \in \Delta_{j}} dP_{i}(\omega^{n}) + 0\\
\nonumber &= \sum_{j \ne i} \sum_{n>0}\int_{\omega^{n} \in \Delta_{j}} \frac{dP_{i}}{dP_{j}}(\omega^{n})dP_{j}(\omega^{n})\\
\label{eqn:proof of propostion prob of wrong detection equation 4} &\le \sum_{j \ne i} \sum_{n>0}\int_{\omega^{n} \in \Delta_{j}} \frac{1}{(M-1)L}dP_{j}(\omega^{n})\\
\nonumber& \le \frac{1}{(M-1)L} \sum_{j \ne i} P_{j}(\Delta_{j})\\
\nonumber & \le \frac{1}{L}.
 \end{align}
The equality in the second step is valid as we have shown in Proposition \ref{prop:finite stopping time} that the stopping time is finite with probability 1. The inequality (\ref{eqn:proof of propostion prob of wrong detection equation 4}) follows because under $H = H_{j}$, $\omega^{n} \in \Delta_{j}$ implies $Z_{ji}(n) \ge \log((M-1)L)$, that is, $\frac{dP_{i}}{dP_{j}}(\omega^{n}) < \frac{1}{(M-1)L}$.
\hfill \IEEEQEDclosed

\subsection{Proof of Theorem \ref{theorem: upper bound on stopping time of SA}: Achievability}
\label{ref:lemma:achievability proof}
We assume (I) and (IIb). All statements in this proof are under $H = H_{i}$ and under {\it Sluggish Procedure A}.  We follow the proof technique of Chernoff \cite[Lem. 2]{ref:195909AMS_Che}. Chernoff's proof technique does not go through completely because unlike in {\it Procedure A}, the next action in {\it Sluggish Procedure A} is not conditionally independent of the previous action, given the current likelihood values. A similar issue was addressed by Nitinawarat and Veeravalli in \cite{2013arXiv1310.1844_NitinawaratVeeravalli}, and we will adapt their proof technique to our setting.

Let us first setup some notation. Fix $\epsilon > 0$. Define $$D_{ij}:= \sum_{a \in \mathcal{A}} \lambda_{i}(a) D(q_{i}^{a} \Vert q_{j}^{a}),$$ where $\lambda_{i}$ is as defined in (\ref{eqn:optimal_lambda}). Let $D_{i}$ be as defined by (\ref{eqn:D_i}), i.e., $D_{i} = \min_{j \ne i}D_{ij}$. Under the {\it Sluggish Procedure A}, the transition probability matrix $TP(\theta({n}))$ of the action process $A_{n}$ at time $n$ is given by 

\begin{align}
\label{eqn:transition_prob_matrix}
 TP(\theta(n)) = (1-\eta) \mathbf{I}+\eta \left(\mathbf{\underbar{1}} \: \lambda_{\theta(n)}^{T}\right).
\end{align}

It is easy to verify that the stationary distribution associated with $TP(\theta({n}))$ is $\lambda_{\theta(n)}$. Define $\mathcal{F}_{k-1} := \sigma(X^{k-1},A^{k-1})$, the $\sigma$-field generated by the random variables $(X^{k-1},A^{k-1})$.

We now upper bound the expected time to make a decision under {\it Sluggish Procedure A} as follows:
\begin{align}
 \nonumber E_{i}\left[\tau(\pi_{SA}(L,\eta))\right] & \le E_{i}\left[\tau(\pi_{SA}^{i}(L,\eta))\right]\\
 \nonumber & = \sum_{n \ge 0} P_{i} \left(\tau(\pi_{SA}^{i}(L,\eta)) > n\right)\\
  \nonumber & \le \frac{(1+\epsilon) \log (L(M-1))}{D_{i}}\\
\label{eqn:bounding expected stopping time equation 3} & \hspace{0.5 cm}+ \sum_{n \ge \tilde{n}} P_{i} \left(\tau(\pi_{SA}^{i}(L,\eta)) > n\right),
 \end{align}
where $$\tilde{n} =  \frac{(1+\epsilon) \log (L(M-1))}{D_{i}}.$$
To complete the proof, we will now show that for any $\epsilon >0$, the second term on the right-hand side of (\ref{eqn:bounding expected stopping time equation 3}) goes to zero as $L \rightarrow \infty$. Let us first analyse  a term in the summation. We claim that each term  decays exponentially with $n$. So the tail sum vanishes as $L \rightarrow \infty$, because $\tilde{n} \rightarrow \infty$. This suffices to complete the proof of Theorem \ref{theorem: upper bound on stopping time of SA}.

We now proceed to prove the claim. Observe that
\begin{align*}
 P_{i} &\left(\tau(\pi_{SA}^{i}(L,\eta)) > n\right)\\
 & \le P_{i} \left(\min_{j \ne i} Z_{ij}(n) \le \log(L(M-1))\right)\\
& \le \sum_{j \ne i}  P_{i} \left(Z_{ij}(n) \le \log(L(M-1))\right).
\end{align*}

Fix one $j \ne i$. (The same analysis holds for other $j$.) Then
{
\begin{align}
 \nonumber P_{i}  & \left(Z_{ij}(n) \le \log(L(M-1))\right)\\
 \nonumber & = P_{i} \left(\sum_{k=1}^{n}\Delta Z_{ij}(k) \le \log(L(M-1))\right) \\
  \nonumber& = P_{i} \left(\sum_{k=1}^{n} \left( \Delta Z_{ij}(k) - E_{i} \left[ \Delta Z_{ij}(k)\vert \mathcal{F}_{k-1}\right] + \epsilon' \right) \right. \\
  \nonumber & \hspace{1.2 cm} + \sum_{k=1}^{n} \left(E_{i} \left[ \Delta Z_{ij}(k)\vert \mathcal{F}_{k-1}\right] - D_{ij} + \epsilon' \right) \\
  \nonumber & \hspace{1.2 cm} +  n \left(D_{ij} - 2\epsilon' \right) \le \log (M-1)L \Bigg) \\
  \nonumber & \le P_{i}\left( \sum_{k=1}^{n} \left( \Delta Z_{ij}(k) - E_{i} \left[ \Delta Z_{ij}(k)\vert \mathcal{F}_{k-1}\right] + \epsilon' \right) < 0 \right) \\
  \nonumber& \hspace{0.7 cm} + P_{i} \left( \sum_{k=1}^{n} \left(E_{i} \left[ \Delta Z_{ij}(k)\vert \mathcal{F}_{k-1}\right] - D_{ij} + \epsilon' \right) < 0 \right) \\
 \label{eqn:log likelihood ratio Z_ij equation 2} &  \hspace{0.7 cm} + P_{i} \left( n (D_{ij} - 2\epsilon') \le \log(L(M-1))  \right).
\end{align}
}Look at the first probability term in (\ref{eqn:log likelihood ratio Z_ij equation 2}). Each entry within the summation has a positive mean and, from Chernoff's bounding technique in \cite[Lem. 2]{ref:195909AMS_Che}, there exists a $b(\epsilon') > 0$ such that 
\begin{align*}
 P_{i} & \left( \sum_{k=1}^{n} \left( \Delta Z_{ij}(k) - E_{i} \left[ \Delta Z_{ij}(k)\vert \mathcal{F}_{k-1}\right] + \epsilon' \right) < 0 \right) \le e^{-n b(\epsilon')}. 
\end{align*}

The third probability term is $0$ if we choose an $\epsilon'$ small enough such that $n(D_{ij} - 2 \epsilon') > \log (L(M-1))$, for all $n > \tilde{n}$. Indeed, any $\epsilon'$ satisfying $0 < \epsilon' < \frac{\epsilon}{1+\epsilon}\frac{D_{i}}{2}$ suffices. So set $\epsilon' = \frac{\epsilon}{1+\epsilon} \frac{D_{i}}{4}$.

We now proceed to show that the second term also decays exponentially to zero. Let $T_{i}$ be as defined in (\ref{eqn:Ti}). For a suitably chosen $\epsilon''$, and we will soon indicate how to choose it, we have
\begin{align*}
 P_{i} & \left( \sum_{k=1}^{n} \left(E_{i} \left[ \Delta Z_{ij}(k)\vert \mathcal{F}_{k-1}\right] - D_{ij} + \epsilon' \right) < 0 \right)  \\
& \le P_{i} \left( \sum_{k=1}^{n} \left(E_{i} \left[ \Delta Z_{ij}(k)\vert \mathcal{F}_{k-1}\right] - D_{ij} + \epsilon' \right) < 0,~ T_{i} \le n\epsilon'' \right)\\
& \hspace{0.7 cm} + P_{i}(T_{i} > n \epsilon'').
\end{align*}
From Lemma \ref{lemma:exponential decay of Ti}, the second probability term on the right-hand side decays exponentially with $n$. To show that the first probability term on the right-hand side decays exponentially with $n$, we use a technique of Nitinawarat and Veeravalli \cite[(6.23)]{2013arXiv1310.1844_NitinawaratVeeravalli}.

First, we indicate how to choose $\epsilon''$. Define 
\begin{align*} 
\tilde{C} & = \min_{a \in \mathcal{A}} E_{i}\left[\Delta Z_{ij}(k)\vert A_{k}=a \right] - D_{ij}\\
& = \min_{a \in \mathcal{A}} D(q_{i}^{a}\Vert q_{j}^{a}) - D_{ij}.
\end{align*}
Since $D_{ij}$ is the $\lambda_{i}$-weighted average of $D(q_{i}^{a}\Vert q_{j}^{a})$, we have $\tilde{C} \le 0$. Choose $\epsilon''$ small enough so that $\tilde{\epsilon} := \epsilon'+\epsilon''\tilde{C} > 0$. We then have
\begin{align*}
 P_{i}  & \left( \sum_{k=1}^{n} \left(E_{i} \left[ \Delta Z_{ij}(k)\vert \mathcal{F}_{k-1}\right] - D_{ij} + \epsilon' \right) < 0,~ T_{i} \le n\epsilon'' \right) \\
 & = P_{i} \left(\sum_{k=1}^{\lfloor n \epsilon'' \rfloor} \left(E_{i} \left[ \Delta Z_{ij}(k)\vert \mathcal{F}_{k-1}\right] - D_{ij} + \epsilon' \right) \right.\\
 & \hspace{1.2 cm}+ \sum_{ k= \lfloor n \epsilon'' \rfloor + 1}^{n} \left(E_{i} \left[ \Delta Z_{ij}(k)\vert \mathcal{F}_{k-1}\right] - D_{ij} + \epsilon' \right)  < 0,\\
 &  \hspace{1.2 cm} T_{i} \le n\epsilon'' \Bigg)\\
 & \le P_{i} \left(\lfloor n \epsilon'' \rfloor (\tilde{C}+\epsilon') \right.\\
 &  \hspace{1.2 cm} + \sum_{k= \lfloor n \epsilon'' \rfloor + 1}^{n} \left(E_{i} \left[ \Delta Z_{ij}(k)\vert \mathcal{F}_{k-1}\right] - D_{ij} + \epsilon' \right)  < 0, \\
 & \hspace{1.2 cm} T_{i} \le n\epsilon'' \Bigg)\\
 & \le P_{i} \left( \sum_{k= \lfloor n \epsilon'' \rfloor + 1}^{n} \left(E_{i} \left[ \Delta Z_{ij}(k)\vert \mathcal{F}_{k-1}\right] - D_{ij} +  \tilde{\epsilon} \right)  < 0, \right. \\
 &  \hspace{1.2 cm} T_{i} \le n\epsilon'' \Bigg)\\
 & \le \tilde{P}_{i} \left( \sum_{k=\lfloor n \epsilon'' \rfloor}^{n} \left(E_{i} \left[ \Delta Z_{ij}(k)\vert \mathcal{F}_{k-1}\right] - D_{ij}+ \tilde{\epsilon} \right) < 0 \right)\\
 & \le C e^{-n\tilde{b}(\tilde{\epsilon})}, \numberthis \label{eqn:exponential decay stationary markov process equation 5}
\end{align*}
for some $C > 0$ and some $\tilde{b}(\tilde{\epsilon}) > 0$. The second inequality follows from the fact that $\tilde{C} \le E_{i} \left[ \Delta Z_{ij}(k)\vert \mathcal{F}_{k-1}\right] - D_{ij}$, for all $k$. The third inequality follows from the choice of $\tilde{\epsilon}$ and the fact that $$\lfloor n \epsilon'' \rfloor (\tilde{C}+\epsilon')+ (n - \lfloor n \epsilon'' \rfloor) \epsilon' \ge (n-\lfloor n \epsilon'' \rfloor) \tilde{\epsilon}.$$ $\tilde{P}_{i}$ is a new measure under which actions are taken according to {\it Sluggish Procedure A} but assuming $\theta(n) = i \quad \forall n$, and the observations are conditionally independent of past observations and actions, given the current action. Consequently, under $\tilde{P}_{i}$, the action process $A_{n}$ is a stationary Markov Chain with transition probability matrix $TP(i)$. By the ergodic theorem and concentration inequalities for Markov Chains \cite{1998_AnnalsAppliedProb_Lezaud}, this term also decays exponentially with $n$, which is (\ref{eqn:exponential decay stationary markov process equation 5}).
\hfill \IEEEQEDclosed



\bibliographystyle{../../IEEEtran/bibtex/IEEEtran}
{
\bibliography{../../IEEEtran/bibtex/IEEEabrv,../../BIB/ISITbib}
}


\end{document}